\documentclass[twocolumn,letterpaper,aps,prl,superscriptaddress,showpacs,floatfix,preprintnumbers,longbibliography]{revtex4-2}

\usepackage{lineno}
\usepackage{graphicx}
\usepackage{amsmath}
\usepackage{hyperref}
\begin{document}

% Should be commented out in the submission
%\preprint{\sl \new{version 1.0} \today. Text: \new{new}, \red{old}, \note{note}}

%Title of paper
\title{Energy Correlators Within Jets in Transversely Polarized Proton-Proton Collisions at $\sqrt{s} = 200$ GeV }

%\author{The STAR Collaboration}
\affiliation{Academia Sinica, Nankang, 115}
\affiliation{Panchayat College (PC), Bargarh, affiliated with Sambalpur University (SU), Odisha 768028, India}
\affiliation{Abilene Christian University, Abilene, Texas   79699}
\affiliation{Alikhanov Institute for Theoretical and Experimental Physics NRC "Kurchatov Institute", Moscow 117218}
\affiliation{Argonne National Laboratory, Argonne, Illinois 60439}
\affiliation{American University in Cairo, New Cairo 11835, Egypt}
\affiliation{Ball State University, Muncie, Indiana, 47306}
\affiliation{Brookhaven National Laboratory, Upton, New York 11973}
\affiliation{University of Calabria \& INFN-Cosenza, Rende 87036, Italy}
\affiliation{University of California, Berkeley, California 94720}
\affiliation{University of California, Davis, California 95616}
\affiliation{University of California, Los Angeles, California 90095}
\affiliation{University of California, Riverside, California 92521}
\affiliation{Central China Normal University, Wuhan, Hubei 430079 }
\affiliation{University of Illinois at Chicago, Chicago, Illinois 60607}
\affiliation{Chongqing University, Chongqing, 401331}
\affiliation{Creighton University, Omaha, Nebraska 68178}
\affiliation{Czech Technical University in Prague, FNSPE, Prague 115 19, Czech Republic}
\affiliation{National Institute of Technology Durgapur, Durgapur - 713209, India}
\affiliation{ELTE E\"otv\"os Lor\'and University, Budapest, Hungary H-1117}
\affiliation{Frankfurt Institute for Advanced Studies FIAS, Frankfurt 60438, Germany}
\affiliation{Fudan University, Shanghai, 200433 }
\affiliation{Guangxi Normal University, Guilin, 541004}
\affiliation{University of Heidelberg, Heidelberg 69120, Germany }
\affiliation{University of Houston, Houston, Texas 77204}
\affiliation{Huzhou University, Huzhou, Zhejiang  313000}
\affiliation{Indian Institute of Science Education and Research (IISER), Berhampur 760010 , India}
\affiliation{Indian Institute of Science Education and Research (IISER) Tirupati, Tirupati 517507, India}
\affiliation{Indian Institute Technology, Patna, Bihar 801106, India}
\affiliation{Indiana University, Bloomington, Indiana 47408}
\affiliation{Institute of Modern Physics, Chinese Academy of Sciences, Lanzhou, Gansu 730000 }
\affiliation{University of Jammu, Jammu 180001, India}
\affiliation{Joint Institute for Nuclear Research, Dubna 141 980}
\affiliation{Kent State University, Kent, Ohio 44242}
\affiliation{University of Kentucky, Lexington, Kentucky 40506-0055}
\affiliation{Lanzhou University, Lanzhou, 730000}
\affiliation{Lawrence Berkeley National Laboratory, Berkeley, California 94720}
\affiliation{Lehigh University, Bethlehem, Pennsylvania 18015}
\affiliation{Lovely Professional University, Jalandhar - Delhi G.T. Road, Pagwara, Panjab, 144411, India}
\affiliation{Max-Planck-Institut f\"ur Physik, Munich 80805, Germany}
\affiliation{Michigan State University, East Lansing, Michigan 48824}
\affiliation{National Research Nuclear University MEPhI, Moscow 115409}
\affiliation{National Institute of Science Education and Research, HBNI, Jatni 752050, India}
\affiliation{National Cheng Kung University, Tainan 70101 }
\affiliation{The Ohio State University, Columbus, Ohio 43210}
\affiliation{Panjab University, Chandigarh 160014, India}
\affiliation{NRC "Kurchatov Institute", Institute of High Energy Physics, Protvino 142281}
\affiliation{Purdue University, West Lafayette, Indiana 47907}
\affiliation{Rice University, Houston, Texas 77251}
\affiliation{Rutgers University, Piscataway, New Jersey 08854}
\affiliation{University of Science and Technology of China, Hefei, Anhui 230026}
\affiliation{South China Normal University, Guangzhou, Guangdong 510631}
\affiliation{Sejong University, Seoul, 05006, Korea, Republic Of}
\affiliation{Shandong University, Qingdao, Shandong 266237}
\affiliation{Shanghai Institute of Applied Physics, Chinese Academy of Sciences, Shanghai 201800}
\affiliation{Southern Connecticut State University, New Haven, Connecticut 06515}
\affiliation{State University of New York, Stony Brook, New York 11794}
\affiliation{Instituto de Alta Investigaci\'on, Universidad de Tarapac\'a, Arica 1000000, Chile}
\affiliation{Temple University, Philadelphia, Pennsylvania 19122}
\affiliation{Texas A\&M University, College Station, Texas 77843}
\affiliation{Texas Southern University, Houston, Texas, 77004}
\affiliation{University of Texas, Austin, Texas 78712}
\affiliation{Tsinghua University, Beijing 100084}
\affiliation{University of Tsukuba, Tsukuba, Ibaraki 305-8571, Japan}
\affiliation{University of Chinese Academy of Sciences, Beijing, 101408}
\affiliation{Valparaiso University, Valparaiso, Indiana 46383}
\affiliation{Variable Energy Cyclotron Centre, Kolkata 700064, India}
\affiliation{Warsaw University of Technology, Warsaw 00-661, Poland}
\affiliation{Wayne State University, Detroit, Michigan 48201}
\affiliation{Wuhan University of Science and Technology, Wuhan, Hubei 430065}
\affiliation{Yale University, New Haven, Connecticut 06520}

\author{B.~E.~Aboona}\affiliation{Texas A\&M University, College Station, Texas 77843}
\author{J.~Adam}\affiliation{Czech Technical University in Prague, FNSPE, Prague 115 19, Czech Republic}
\author{G.~Agakishiev}\affiliation{Joint Institute for Nuclear Research, Dubna 141 980}
\author{I.~Aggarwal}\affiliation{Panjab University, Chandigarh 160014, India}
\author{M.~M.~Aggarwal}\affiliation{Panjab University, Chandigarh 160014, India}
\author{Z.~Ahammed}\affiliation{Variable Energy Cyclotron Centre, Kolkata 700064, India}
\author{E.~C.~Aschenauer}\affiliation{Brookhaven National Laboratory, Upton, New York 11973}
\author{A.~Aitbayev}\affiliation{Joint Institute for Nuclear Research, Dubna 141 980}
\author{I.~Alekseev}\affiliation{Alikhanov Institute for Theoretical and Experimental Physics NRC "Kurchatov Institute", Moscow 117218}\affiliation{National Research Nuclear University MEPhI, Moscow 115409}
\author{E.~Alpatov}\affiliation{National Research Nuclear University MEPhI, Moscow 115409}
% ...

\collaboration{STAR Collaboration}\noaffiliation

\date{\today}

\begin{abstract}
We report the first measurement of one- and two-point energy correlators within jets in transversely polarized proton-proton collisions at $\sqrt{s}=200$ GeV, using the STAR detector at RHIC. These observables quantify the energy-weighted angular distribution of single hadrons and hadron pairs within jets, respectively. Sizable spin-dependent asymmetries are observed for $\pi^+$, $\pi^-$, and $\pi^+\pi^-$ pairs, revealing the onset of nonperturbative dynamics at specific angular scales. By projecting the fragmentation dynamics onto Mellin moments, these measurements provide sensitivity to the nucleon's transversity while minimizing uncertainties from nonperturbative fragmentation functions. These results establish energy correlators as a novel and precise probe of nucleon structure and open a promising avenue for three-dimensional nucleon tomography at the future Electron-Ion Collider.
\end{abstract}
\maketitle

A fundamental goal of quantum chromodynamics (QCD) is to unravel the multidimensional structure of the nucleon and understand how its energetic constituents, quarks and gluons, evolve into observable hadrons. This is particularly challenging to study in the nonperturbative regime, where confinement and spin-momentum correlations play a dominant role. Energy-energy correlators (EECs) offer a powerful framework to address this challenge. Originally introduced in electron-positron annihilation~\cite{PhysRevLett.41.1585}, EECs are infrared- and collinear-safe jet substructure observables that quantify the angular correlation of energy flow within a jet~\cite{Liu:2022wop}. By directly measuring energy flow rather than individual hadrons, they provide a unique probe of the transition from the perturbative parton shower to nonperturbative hadronization, bridging the regimes of asymptotic freedom and color confinement~\cite{moult2025energycorrelatorsjourneytheory}.\par

Recent measurements from unpolarized proton-proton collisions at the Large Hadron Collider (LHC) by CMS~\cite{CMS:2024mlf} and ALICE~\cite{ALICE:2024dfl}, and at the Relativistic Heavy Ion Collider (RHIC) by STAR~\cite{STAR:2025jut}, have demonstrated the versatility of EECs in connecting theory and experiment across a broad range of collision and jet energies. These results reveal a distinct transition from perturbative to nonperturbative dynamics consistent with a universal hadronization scale, occurring at an angular scale that decreases as jet transverse momentum increases~\cite{Komiske:2022enw}. The energy-weighted angular distributions follow expected QCD trends: asymptotic freedom at large angles, confinement at small angles, and a transition region in between associated with the hadronization process.\par

Extending EEC measurements to transversely polarized proton-proton collisions provides a novel probe of spin-dependent fragmentation dynamics within jets, offering new sensitivity to transverse-momentum-dependent (TMD) physics and laying the groundwork for future studies at the Electron-Ion Collider~\cite{Kang:2023big, Song:2025bdj}. Perturbative QCD predicts spin-dependent effects from hard scattering to be on the order of $10^{-4}$~\cite{PhysRevLett.41.1689}, so any sizable azimuthal asymmetry observed in the correlator is a direct signature of nonperturbative dynamics. Uniquely, the EEC provides access to the angular dependence of spin-dependent correlations, allowing us to spatially map the onset of spin-dependent hadronization and locate the specific angular scale where the dynamics transition from the perturbative regime to the formation of the final-state hadron. Furthermore, by projecting correlated energy flow onto Mellin moments~\cite{liu2024tmdssemiinclusiveenergycorrelators, 1lz2-3fm9, kang2026simplifiedapproachextractingnucleon,kang2026transversespindependentenergyenergycorrelators}, defined as energy-weighted integrals of the fragmentation functions, EECs minimize the modeling uncertainties inherent in standard fragmentation functions. This establishes a robust pathway toward extracting the poorly constrained chiral-odd transversity distribution, a fundamental pillar of nucleon structure, complementing previous STAR results on hadron-in-jet and di-hadron observables~\cite{STAR:2015jkc, STAR:2022hqg}.\par

\begin{figure*}
	\centering
	\includegraphics[width=1.0\linewidth]{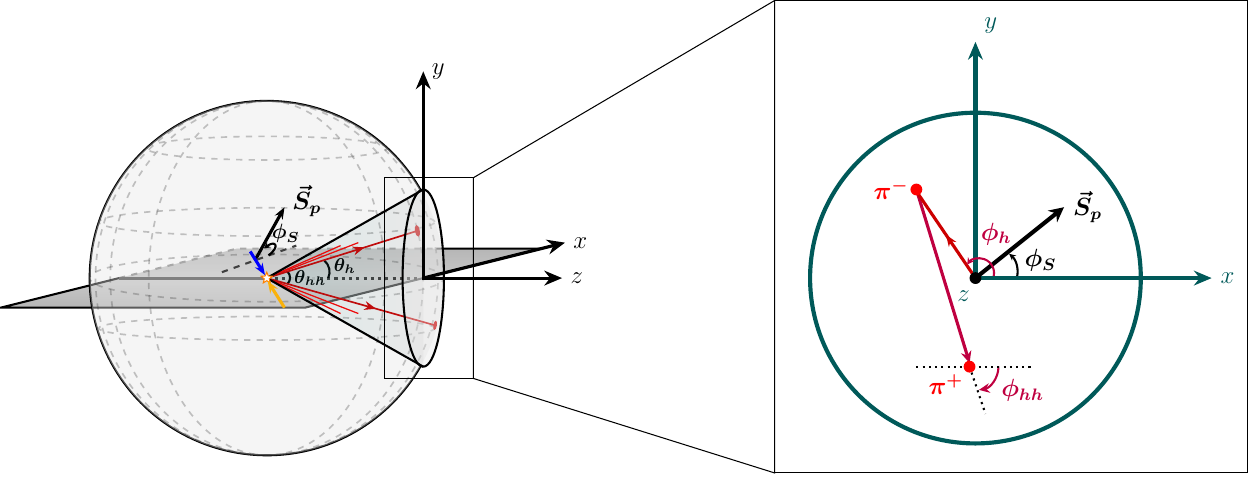}
	\caption{Representation of the jet scattering plane, the transverse spin vector of the polarized proton beam, $\vec{S}_p$, the angular distances $\theta_h$ and $\theta_{hh}$, and the azimuthal angles $\phi_S$, $\phi_h$, and $\phi_{hh}$. The azimuthal angles are defined in the plane transverse to the jet axis, which lie along the $z$ direction. The polarized proton beam lies in the jet scattering plane (the $x$-$z$ plane), and $\phi_S$ denotes the azimuthal angle of $\vec{S}_p$. The angle $\phi_{hh}$ is defined by the pair vector pointing from the negatively charged hadron to the positively charged hadron. The blue and yellow arrows indicate the directions of the incoming proton beams.}
	\label{fig:Phi_H}
\end{figure*}

In this work, we present the first measurement of one- and two-point energy correlators within jets in transversely polarized proton-proton collisions at $\sqrt{s} = 200$ GeV. The data were collected by the STAR experiment in 2015, corresponding to an integrated luminosity of 52~pb$^{-1}$ with an average beam polarization of $\langle P \rangle = 57\%$ (scale uncertainty of $\Delta P/\langle P \rangle = 3.0\%$)~\cite{ref:RHICPolG}. The primary detector subsystems used in this analysis include the Time Projection Chamber (TPC)~\cite{ref:tpcNIM}, the Barrel and Endcap Electromagnetic Calorimeters~\cite{ref:bemcNIM,ref:eemcNIM}, and the Time-of-Flight (TOF) detector~\cite{ref:starTOF, Chen:2024aom}. Events were selected using jet patch (JP) triggers requiring transverse electromagnetic energy deposits above specific thresholds within a $1.0 \times 1.0$ region in $\eta$-$\phi$.\par

We take the jet axis to define the $z$ direction and the jet scattering plane to define the $x$-$z$ plane. The transverse spin vector of the incident polarized proton is denoted by $\vec{S}_p$, and its azimuthal angle in the transverse plane is $\phi_S$, as illustrated in Fig.~\ref{fig:Phi_H}. The one-point energy correlator, defined in Eq.~\ref{Equ:OneEEC}, measures the energy-weighted distribution of final-state hadrons at an angular distance of $\theta_h$ from the jet axis and azimuth $\phi_{h}$ in the transverse plane, as illustrated in Fig.~\ref{fig:Phi_H}. Each hadron inside a jet is weighted by its momentum relative to the jet momentum (momentum fraction, $z_{h}$), and the sum runs over all hadrons in the jet:
\begin{equation}
\Sigma_{\mathrm{EC}}(\theta_h,\phi_h)
=\frac{d\left(\sum_{\mathrm{jet}}\sum_{i \in \mathrm{jet}} z_{h,i}\right)}{d\theta_h d\phi_h}.
\label{Equ:OneEEC}
\end{equation}

The two-point energy-energy correlator in Eq.~\ref{Equ:TwoEEC} measures the energy-weighted distribution of hadron pairs as a function of their opening angle $\theta_{hh}$ and the relative azimuth $\phi_{hh}$ between hadrons $i$ and $j$ within a jet. The weights are the product of the hadron momentum fractions $z_{h,i} z_{h,j}$. To probe charge-dependent spin correlations, the sum runs over all $h^+h^-$ pairs in the jet, and $\phi_{hh}$ is defined by the pair vector pointing from the negatively charged hadron to the positively charged one with respect to the $x$-axis:
\begin{equation}
\Sigma_{\mathrm{EEC}} (\theta_{hh}, \phi_{hh}) = \frac{d\left(\sum_{\text{jet}} \sum_{i,j \in \mathrm{jet}} z_{h,i} z_{h,j}\right)}{d\theta_{hh} d\phi_{hh}}.
\label{Equ:TwoEEC}
\end{equation}

These distributions map the angular structure of correlated energy flow onto perturbative parton splittings at large $\theta_{hh}$ and nonperturbative hadronization at small $\theta_{hh}$. In transversely polarized collisions, their azimuthal dependence relative to $\phi_S$ reveals spin-dependent modulations. Modulations of the form $\sin(\phi_S - \phi_{h})$ and $\sin(\phi_S - \phi_{hh})$ provide direct sensitivity to the coupling between the nucleon's transversity and spin-dependent fragmentation, such as the Collins effect~\cite{Collins:1992kk} and interference fragmentation functions~\cite{Jaffe:1997hf}.\par

Jet and hadron selections follow those of previous STAR measurements at 200~GeV~\cite{STAR:2022hqg}. The inputs to the jet finder are the charged particle momenta measured by the TPC and the neutral-energy depositions observed by the calorimeter towers, corrected for the energy deposited by matched charged tracks. Jets are reconstructed using the anti-$k_\mathrm{T}$ algorithm~\cite{ref:AntiKtJets} implemented in the FastJet 3.0.6 package~\cite{ref:FastJet} with a radius of $R = 0.6$. Jets included in this analysis are required to have a pseudorapidity $\eta_{\mathrm{jet}}$ spanning $0 < \eta_{\mathrm{jet}} < 0.9$ relative to the polarized proton beam. To minimize the effects of poor momentum resolution and charge discrimination, jets containing tracks with $p_{\mathrm{T}} > 20$~GeV/$c$ are excluded. Constituent hadrons are restricted to momentum fractions $0.1 < z_h < 0.8$, suppressing the underlying event and further reducing sensitivity to high-momentum resolution limits. For the one-point energy correlator, a minimum separation $\Delta R_h = \sqrt{(\eta_{\mathrm{jet}}-\eta_{h})^2 + (\phi_{\mathrm{jet}}-\phi_{h})^2} > 0.02$ between the hadron and the jet axis is imposed to ensure a reliable reconstruction of $\phi_h$. For the two-point correlator, constituent hadrons within the same jet are paired inclusively. To minimize trigger bias near the hardware thresholds, every jet must geometrically match a triggering jet patch and satisfy a minimum $p_{\mathrm{T,jet}}$ of $6.0$~GeV/$c$ (for JP1) or $8.4$~GeV/$c$ (for JP2).\par

Charged pions are identified via their observed energy loss $dE/dx_{obs}$ in the TPC. We define $n_\sigma(\pi)$ as the number of standard deviations from the expected pion value: 
\begin{equation}
n_\sigma(\pi) \equiv \frac{1}{\sigma_{exp}}\mathrm{ln}\left(\frac{dE/dx_{obs}}{dE/dx_{\pi,calc}}\right),
\end{equation}
where $dE/dx_{\pi,calc}$ is the expected energy loss for charged pions from the Bichsel formalism~\cite{ref:Bichsel:2006cs}, and $\sigma_{exp}$ is the detector resolution~\cite{ref:PeakShift,ref:ShaoParticleId}. The value of $n_\sigma(\pi)$ is required to be within (-1, 2) in order to remove a large fraction of contamination from kaons, protons, and electrons.\par

The transverse single-spin asymmetries for both the one- and two-point energy correlators are extracted using the cross-ratio method for which both acceptance and luminosity effects cancel to first order~\cite{ref:OhlsenCR}. For a given azimuthal bin, we combine weighted yields from detector halves opposite in azimuth when the spin orientations are flipped:
\begin{widetext}
\begin{equation}
\label{equ:cross-ratio}
A_{\mathrm{UT}}^{\sin(\phi_{S}-\phi_{h})}\sin(\phi_{S}-\phi_{h}) \\
= \frac{1}{P}
\frac{\sqrt{\Sigma_{EC}^{\uparrow}[\phi_{S}-\phi_{h}]\Sigma_{EC}^{\downarrow}[(\phi_{S}+\pi)-\phi_{h}]}-\sqrt{\Sigma_{EC}^{\downarrow}[\phi_{S}-\phi_{h}]\Sigma_{EC}^{\uparrow}[(\phi_{S}+\pi)-\phi_{h}]}}{\sqrt{\Sigma_{EC}^{\uparrow}[\phi_{S}-\phi_{h}]\Sigma_{EC}^{\downarrow}[(\phi_{S}+\pi)-\phi_{h}]}+\sqrt{\Sigma_{EC}^{\downarrow}[\phi_{S}-\phi_{h}]\Sigma_{EC}^{\uparrow}[(\phi_{S}+\pi)-\phi_{h}]}},
\end{equation}
\end{widetext}
where $\Sigma_{EC}^{\uparrow}$ ($\Sigma_{EC}^{\downarrow}$) is the one-point energy correlator distribution when the proton beam spin is ``up" (``down"), and $P$ is the beam polarization. An equivalent formalism is applied to the two-point analysis by substituting $\Sigma_{EC}^{\uparrow}$ ($\Sigma_{EC}^{\downarrow}$) with $\Sigma_{EEC}^{\uparrow}$ ($\Sigma_{EEC}^{\downarrow}$), and $\phi_{h}$ to $\phi_{hh}$. In Eq.~\ref{equ:cross-ratio}, $\phi_{S}$ spans the range of $-\pi/2$ to $\pi/2$, while $\phi_{h}$ and $\phi_{hh}$ covers the full azimuth.\par

\begin{figure}[htbp]
	\centering
	\includegraphics[width=1.0\linewidth]{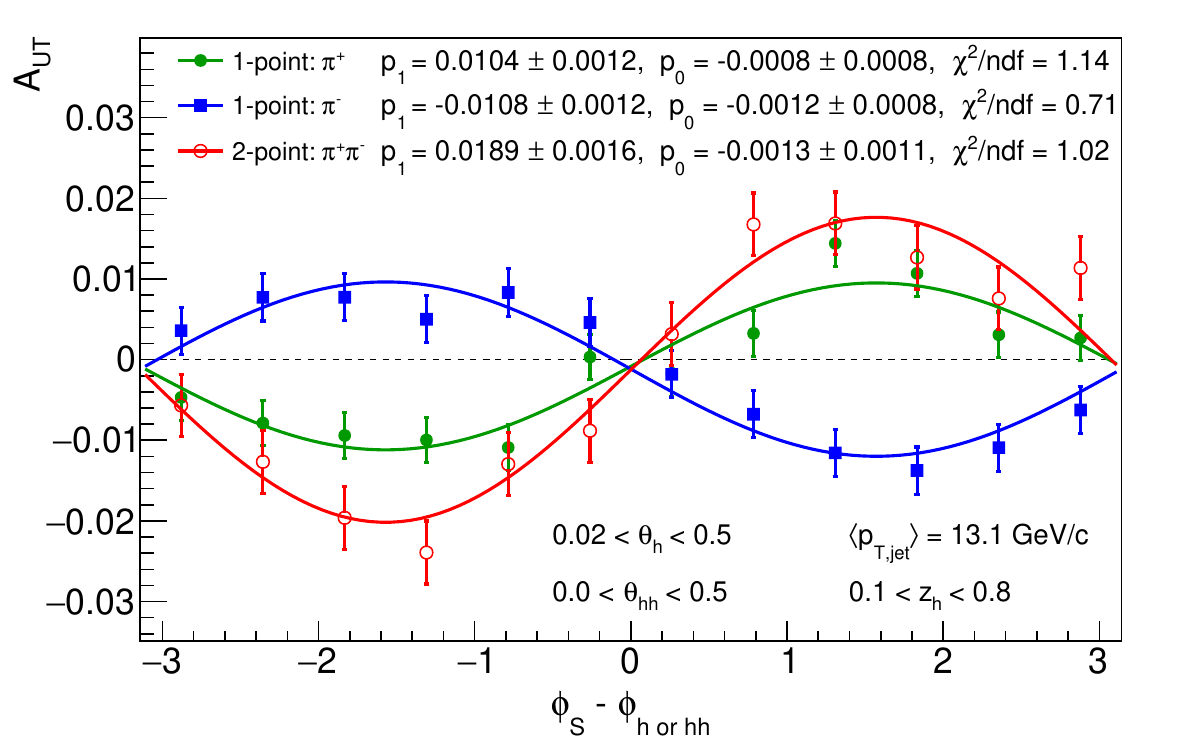}
	\caption{Azimuthal single spin asymmetries as functions of $\phi_S - \phi_{h/hh}$ with average jet transverse momentum of $p_{\mathrm{T,jet}}\sim13.1~\mathrm{GeV}/c$. Error bars represent statistical uncertainties.}
	\label{fig:Fig2}
\end{figure}

As shown in Fig.~\ref{fig:Fig2}, the asymmetries are extracted by binning Eq.~\ref{equ:cross-ratio} into twelve azimuthal bins spanning $(-\pi, \pi)$ and fitting with the sinusoidal function:
\begin{equation}
  \label{equ:asym_fit}
  p_{0} + p_{1} \times \sin(\phi_S-\phi_{h/hh}),
\end{equation}
where $p_{1}$ is the asymmetry, $p_{0}$ is the offset, and $\phi_{h/hh}$ denotes $\phi_h$ for single hadrons and $\phi_{hh}$ for hadron pairs. Asymmetries for the modulations $\sin(\phi_{S}+\phi_{h/hh})$, $\sin(\phi_{S}-2\phi_{h/hh})$, and $\sin(\phi_{S}+2\phi_{h/hh})$ were also examined as a check for harmonic leakage and found to be consistent with zero within statistical uncertainties. Systematic biases arising from non-uniform detector acceptance~\cite{STAR:2022hqg} are estimated to be less than 5\% of the statistical uncertainty and are included in the final systematics.\par

To account for hadron misidentification, the measured asymmetries are unfolded using a purity correction matrix following the same method as Ref.~\cite{STAR:2022hqg}. For the one-point energy correlator, we define non-overlapping samples enriched in pions, kaons, and protons based on their $n_\sigma(\pi)$. For example, for tracks with $0.13 < z_{h} < 0.16$ in a jet with $8.4 < p_{\mathrm{T,jet}} < 9.9~\mathrm{GeV}/c$, $-1<n_\sigma(\pi)<2$, $-5<n_\sigma(\pi)<-1$, and $2<n_\sigma(\pi)<4$ are used for pions, kaons, and protons, respectively. The raw sample asymmetries, $A_{\mathrm{UT}}$, relate to the true species asymmetries, $A_{\mathrm{UT}}^{\mathrm{pure}}$, via the linear transformation by
\begin{equation}
A_{\mathrm{UT}} = A_{\mathrm{UT}}^{\mathrm{pure}} M,
\end{equation}
where $M$ is the $3\times3$ matrix with elements $f_{i}^j$ denoting the fraction of true species $j$ present in sample $i$, extracted from TPC and TOF data~\cite{STAR:2022hqg}. The true asymmetries are obtained by matrix inversion: $A_{\mathrm{UT}}^{\mathrm{pure}} = A_{\mathrm{UT}} M^{-1}$. The pion purity exceeds $90\%$ over the measured kinematic range (See Supplemental Material~\cite{supply}).\par

This formalism is extended to the two-point EEC by applying the single-track selection criteria to define four two-particle samples, labeled $1$ to $4$, which are enriched in the charge-ordered pairs: $\pi^{+}\pi^{-}$, $\pi^{+}K^{-}$, $K^{+}\pi^{-}$, and $K^{+}K^{-}$, respectively. The unfolding utilizes a $4\times4$ mixing matrix, $M$:
\begin{equation}
\label{Equ:Matrix_frac_two}
  M = \\
  \begin{pmatrix}
    f^{\pi^{+}\pi^{-}}_{1} & f^{\pi^{+}\pi^{-}}_{2} & f^{\pi^{+}\pi^{-}}_{3} & f^{\pi^{+}\pi^{-}}_{4}\\
    f^{\pi^{+}K^{-}}_{1}   & f^{\pi^{+}K^{-}}_{2}   & f^{\pi^{+}K^{-}}_{3}   & f^{\pi^{+}K^{-}}_{4}\\
    f^{K^{+}\pi^{-}}_{1}   & f^{K^{+}\pi^{-}}_{2}   & f^{K^{+}\pi^{-}}_{3}   & f^{K^{+}\pi^{-}}_{4}\\
    f^{K^{+}K^{-}}_{1}     & f^{K^{+}K^{-}}_{2}     & f^{K^{+}K^{-}}_{3}     & f^{K^{+}K^{-}}_{4}\\    
  \end{pmatrix},
\end{equation}
relating the raw sample asymmetries, $A_{\mathrm{UT}} = (A^{1}_{\mathrm{UT}}, A^{2}_{\mathrm{UT}}, A^{3}_{\mathrm{UT}}, A^{4}_{\mathrm{UT}})$, to the true pair asymmetries $A_{\mathrm{UT}}^{\mathrm{pure}} = (A_{\mathrm{UT}}^{\pi^{+}\pi^{-}}, A_{\mathrm{UT}}^{\pi^{+}K^{-}}, A_{\mathrm{UT}}^{K^{+}\pi^{-}}, A_{\mathrm{UT}}^{K^{+}K^{-}})$. Pairs containing at least one proton are assigned zero asymmetry, consistent with data showing that the pion-proton asymmetries are statistically consistent with zero within 1$\sigma$. The contribution of these proton-containing pairs is accounted for as a dilution background. The effective $\pi^{+}\pi^{-}$ purity ranges from $80\%$ to $86\%$ (See Supplemental Material~\cite{supply}).

Corrections for kinematic shifts and detector resolution are derived from simulated events generated with \textsc{Pythia} 6.4.28~\cite{ref:Pythia6} using the STAR-tuned Perugia 2012 settings (PARP(90)=0.213)~\cite{ref:PerugiaTunes}. These events are processed through full \textsc{Geant}~3~\cite{ref:Geant} detector simulations and embedded into real zero-bias data to account for pileup and background effects. Detector-level jets are reconstructed using the same selection cuts as real data, while particle-level jets are formed from stable final-state particles. Detector- and particle-level jets are geometrically matched if $\Delta R = \sqrt{\Delta \eta^2 + \Delta \phi^2} < 0.5$, where $\Delta \eta$ and $\Delta \phi$ denote the differences between the jet axes. The probability that a detector jet has a matching particle jet is over 99\%. Hadrons within the selected jets are further matched by charge, with the separation between the matched particle and track satisfy $\Delta R < 0.04$. The measured $p_{\mathrm{T,jet}}$, $\theta_h$, and $\theta_{hh}$ are corrected to particle level using the mean detector-to-particle shifts from simulation, which are approximately $0.3~\mathrm{GeV}/c$, $10^{-3}$, and $10^{-5}$, respectively. Finite azimuthal resolution smears the angle $\phi_{S}-\phi_{h/hh}$, reducing the observed asymmetry. We quantify this effect with a dilution factor, $D$, determined from the simulated detector- and particle-level azimuthal angle distributions, and then corrected by $A_{\mathrm{UT}}/D$. For the one-point result at $p_{\mathrm{T,jet}}\sim13~\mathrm{GeV}/c$, $D$ increases from approximately $0.72$ at the smallest angular separations to above $0.95$ at larger $\theta_{h}$. For the two-point EEC, $D>0.98$ in all bins, corresponding to a correction below $2\%$ (See Supplemental Material~\cite{supply}). The uncertainties associated with these corrections are included in the systematic uncertainty, with their quadrature sum at most $10\%$ of the statistical uncertainty on $A_{\mathrm{UT}}$.\par

Diffuse underlying event (UE) backgrounds from multi-parton and soft beam-remnant interactions can contaminate jet constituents. Evaluated via the off-axis-cone method~\cite{STAR:2022hqg}, UE-related asymmetries were all found to be consistent with zero and are thus treated purely as asymmetry dilution. Systematic uncertainties account for momentum fraction ($z_h$) smearing, selection criteria, trigger bias and UE contamination. The impact of $z_h$ resolution is assessed via a 10\% Gaussian smearing, with the resulting asymmetry shift evaluated via the Barlow test~\cite{Barlow:2002yb}. The dominant systematic uncertainty arises from the $z_h$ selection criteria and is estimated by varying the nominal range ($0.1 < z_h < 0.8$) by $\pm10\%$. Furthermore, because this analysis probes quark transversity, the expected asymmetry is directly proportional to the fraction of quark-initiated jets. The STAR JP trigger can alter this fraction by preferentially selecting specific subprocesses. This trigger bias is evaluated in simulation by calculating the relative change in the quark jet fraction between the unbiased particle level and the detector level; this relative difference is then multiplied by the measured asymmetry (or statistical uncertainty, if larger) as a systematic uncertainty.\par

\begin{figure}[htbp]
	\centering
	\includegraphics[width=1.0\linewidth]{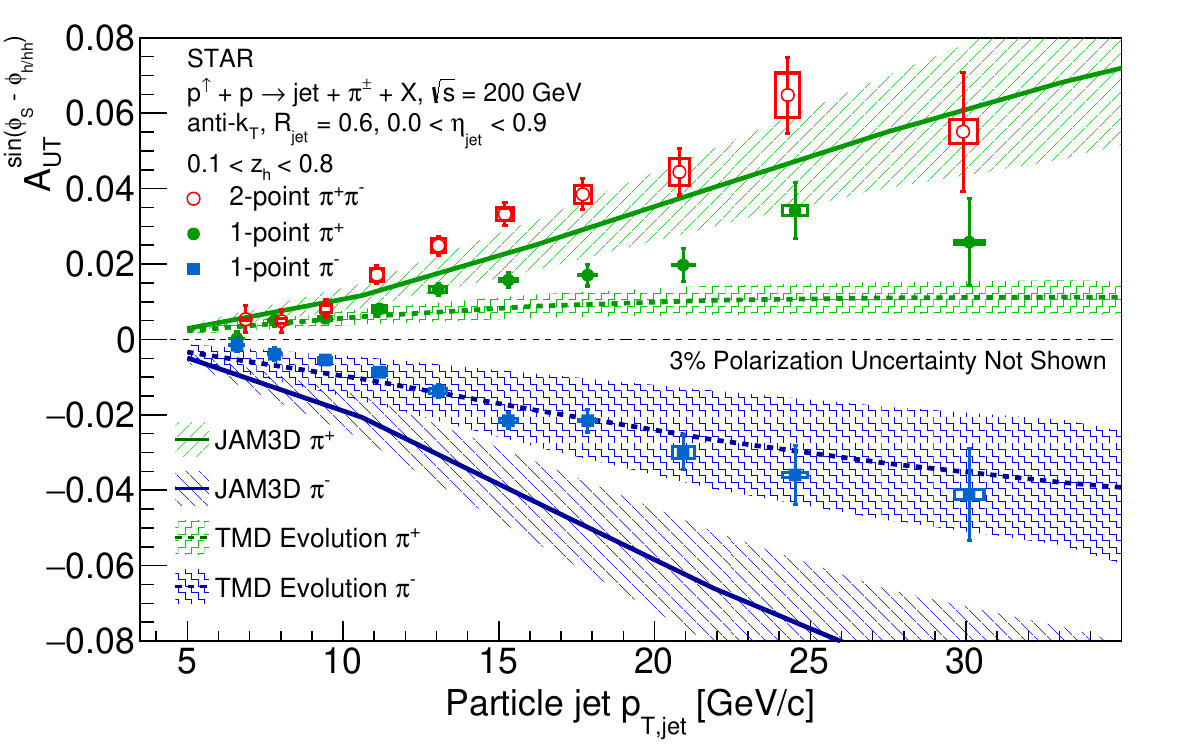}
    \caption{Transverse single spin asymmetries for charged pions inside jets (with $0.02 < \theta_{h} < 0.5$ and $0.0 < \theta_{hh} < 0.5$) as functions of particle jet $p_{\mathrm{T,jet}}$. Markers represent the one-point $\pi^{+}$ (green solid circles) and $\pi^{-}$ (blue solid squares), and the two-point $\pi^{+}\pi^{-}$ (red open circles) correlations. The boxes show the size of the systematic uncertainty. A 3\% vertical scale uncertainty from beam polarization is not shown. The four theory curves shown are discussed in the text.}
	\label{fig:Fig3}
\end{figure}

\begin{figure*}
	\centering
	\includegraphics[width=1.0\linewidth]{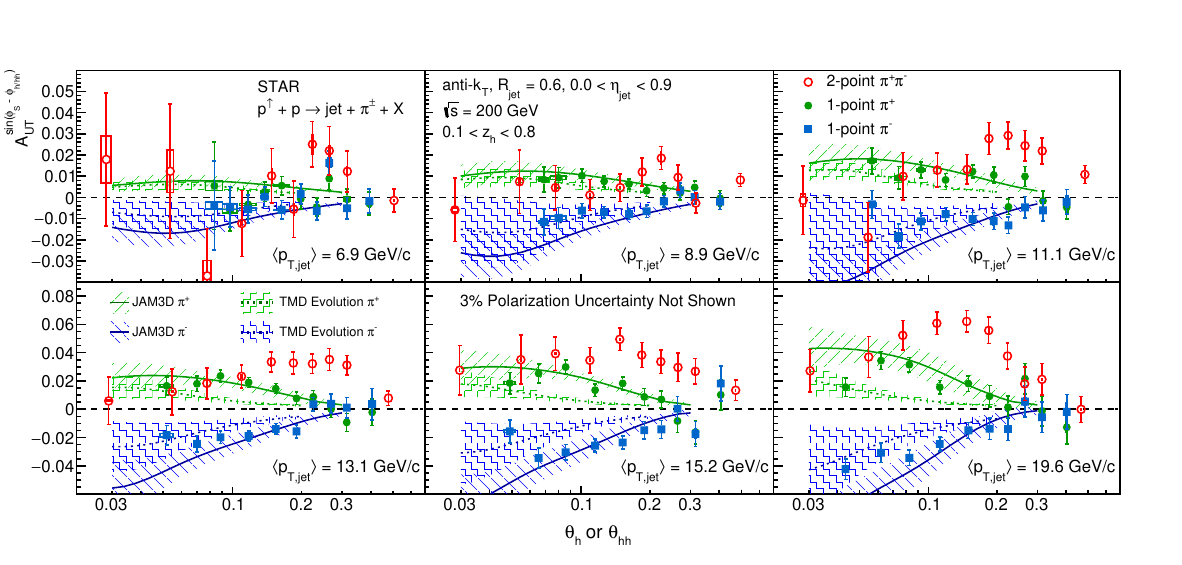}
	\caption{The transverse single spin asymmetry as a function of angular separation $\theta_{h}$ (one-point) or $\theta_{hh}$ (two-point) across six $p_{\mathrm{T,jet}}$ bins. Marker styles follow the convention in Fig.~\ref{fig:Fig3}. Theoretical predictions for one-point results are shown for the TMD evolution framework (dashed lines) and the JAM3D global analysis (solid lines)~\cite{1lz2-3fm9}. Vertical bars represent statistical uncertainties; box heights represent systematic uncertainties, and box widths indicate the angular resolution.}
	\label{fig:Fig4}
\end{figure*}

The final asymmetries for charged pions inside jets, within $0.02 < \theta_{h} < 0.5$ and $0.0 < \theta_{hh} < 0.5$, are presented as functions of jet transverse momentum, $p_{\mathrm{T,jet}}$, in Fig.~\ref{fig:Fig3}. Systematic uncertainties are indicated as open boxes, representing the quadrature sum of contributions from $z_h$ smearing, $z_h$ range selection, underlying event, particle identification, trigger bias, and azimuthal dilution effects; the box widths reflect the jet energy scale uncertainty. Clear nonzero asymmetries are observed, with opposite signs for the one-point $\pi^{+}$ and $\pi^{-}$, and magnitudes that increase with $p_{\mathrm{T,jet}}$. The two-point EEC has similar asymmetry as the one-point $\pi^{+}$ result at low $p_{\mathrm{T,jet}}$, but grows more rapidly with increasing $p_{\mathrm{T,jet}}$, reaching up to $\sim 6\%$ in the highest-$p_{\mathrm{T,jet}}$ bin and about twice the one-point $\pi^{+}$ asymmetry. Theoretical predictions for the one-point energy correlator, adapted to the kinematics of this analysis, are presented using two different frameworks: the full TMD evolution formalism and the JAM3D global QCD analysis, which omits TMD evolution~\cite{1lz2-3fm9}. While the data qualitatively align better with the TMD evolution trends, the observed asymmetry differences between $\pi^+$ and $\pi^-$ are smaller than predicted, suggesting the need for refinements in the flavor dependence of the theoretical framework at these energy scales.\par

Figure~\ref{fig:Fig4} presents the final asymmetries as functions of angular separation ($\theta_{h}$ or $\theta_{hh}$) across six different $p_{\mathrm{T,jet}}$ bins. For the one-point energy correlator, the asymmetries at low $p_{\mathrm{T,jet}}$ ($\leq 11.1~\mathrm{GeV}/c$) are approximately 1\% and vanish for $\theta_{h} > 0.3$. At higher $p_{\mathrm{T,jet}}$ ($\geq 13.1~\mathrm{GeV}/c$), a peak in the asymmetry is observed, which shifts with $p_{\mathrm{T,jet}}$, eventually decreasing to zero at $\theta_{h} \sim 0.2$. The same theoretical predictions are also presented. Here, the data favor the TMD Evolution model at low $p_{\mathrm{T,jet}}$ and small $\theta_{h}$, while the JAM3D framework provides a better description at higher values.\par

For the two-point EEC, the asymmetries exhibit an initial rise followed by a decline with increasing $\theta_{hh}$, with the region of nonzero asymmetry broadening at higher $p_{\mathrm{T,jet}}$. For reference, the transition scale between perturbative and nonperturbative regimes extracted from the unpolarized EEC measurement, $\langle p_{\mathrm{T,jet}}\rangle\theta_{hh}=2.80\pm0.49~\mathrm{GeV}/c$~\cite{STAR:2025jut}, corresponds to $\theta_{hh}^{\rm ref}=0.40\pm0.07$, $0.31\pm0.06$, $0.25\pm0.04$, $0.21\pm0.04$, $0.18\pm0.03$, and $0.14\pm0.03$ in the six $p_{\mathrm{T,jet}}$ bins of Fig.~\ref{fig:Fig4}, respectively. The angular region of nonzero two-point spin asymmetry overlaps with these scale for $\langle p_{\mathrm{T,jet}}\rangle \ge 11.1~\mathrm{GeV}/c$. Empirical fits to the two-point asymmetries (summarized in Supplemental Material~\cite{supply}) reveal that the fitted asymmetry peaks are smaller than the reference angles. Also, the three highest $p_{\mathrm{T,jet}}$ intervals exhibit broad regions of nonzero asymmetry over $\theta_{hh}\approx0.15$--$0.30$, $0.05$--$0.20$, and $0.08$--$0.20$, with asymmetries of approximately $3\%$, $4\%$, and $6\%$, respectively. The observation of sizable asymmetries up to $\theta_{hh}\sim 0.3$ suggests that spin correlations remain relevant within the angular region typically associated with the perturbative parton shower. This provides an intriguing hint that the onset of nonperturbative spin dynamics might manifest at larger angular scales than the confinement transition governing unpolarized bulk energy flow.\par

In summary, we report the first measurement of one- and two-point energy correlators within jets in transversely polarized proton-proton collisions at $\sqrt{s} = 200$ GeV using the STAR detector at RHIC. The experimental precision is comparable to, and in several kinematic regions exceeds, current theoretical uncertainties, providing new quantitative constraints on the phenomenological description of transverse-spin dynamics. These results demonstrate that energy correlators can be effectively extended to polarized hadronic collisions as a theoretically clean and complementary channel for nucleon spin structure studies. This work establishes a foundational baseline for future measurements at the Electron-Ion Collider, where energy correlators will offer a high-precision path toward complete three-dimensional nucleon tomography.\par

We thank Fanyi Zhao for her insightful talk at SPIN2023, which inspired this work, and Ding Yu Shao for the initial discussion and confirming the feasibility of the measurement. We also thank Mei-Sen Gao, Zhong-Bo Kang, Wanchen Li, and Ding Yu Shao for providing their theoretical calculations on the one-point EC results. 

We thank the RHIC Operations Group and SCDF at BNL, the NERSC Center at LBNL, and the Open Science Grid consortium for providing resources and support.  This work was supported in part by the Office of Nuclear Physics within the U.S. DOE Office of Science, the U.S. National Science Foundation, National Natural Science Foundation of China, Chinese Academy of Science, the Ministry of Science and Technology of China and the Chinese Ministry of Education, NSTC Taipei, the National Research Foundation of Korea, Czech Science Foundation and Ministry of Education, Youth and Sports of the Czech Republic, Hungarian National Research, Development and Innovation Office, New National Excellency Programme of the Hungarian Ministry of Human Capacities, Department of Atomic Energy and Department of Science and Technology of the Government of India, the National Science Centre and WUT ID-UB of Poland, German Bundesministerium f\"ur Bildung, Wissenschaft, Forschung and Technologie (BMBF), Helmholtz Association, Ministry of Education, Culture, Sports, Science, and Technology (MEXT), Japan Society for the Promotion of Science (JSPS), and Agencia Nacional de Investigacion y Desarrollo de Chile (ANID), Chile.

Data availability: The data that support the findings of this Letter are openly available~\cite{hepdata.181828}.

% BibTeX:
\bibliography{refs}

\appendix

\section{Supplemental Material}
\setcounter{figure}{0}
\renewcommand{\thefigure}{S\arabic{figure}}

\begin{figure}[htbp]
	\centering
	\includegraphics[width=1.0\linewidth]{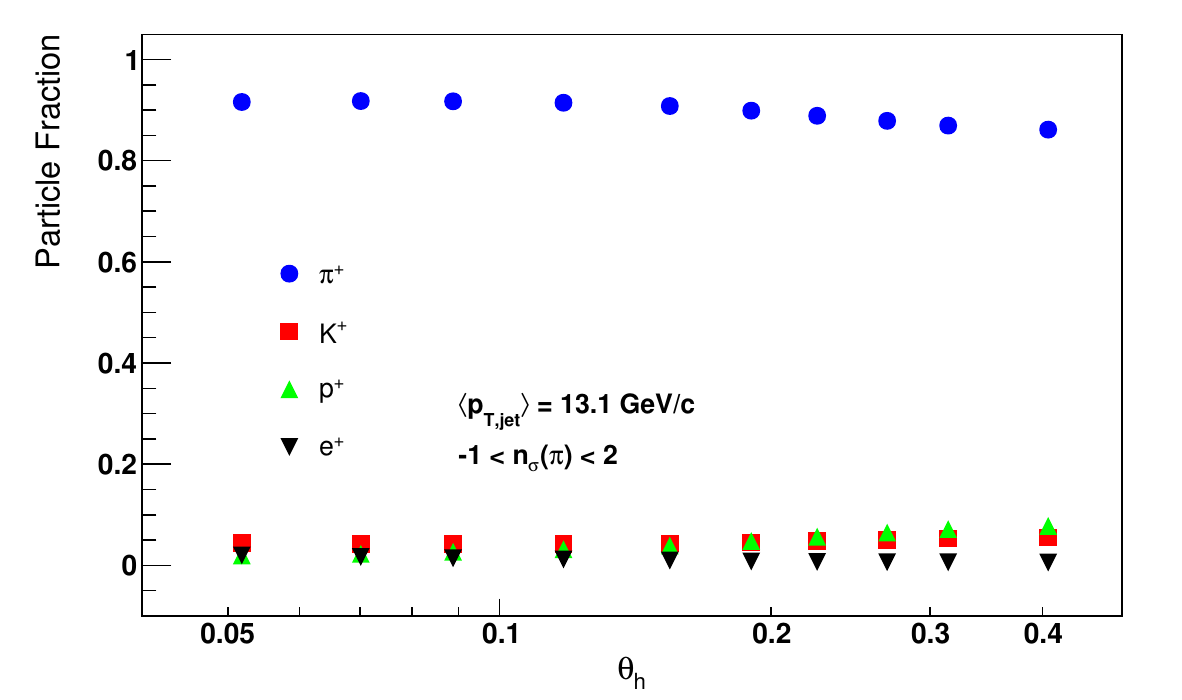}
    \caption{Fractions of charged-particle species as a function of $\theta_h$ for tracks satisfying the nominal pion-enriched selection, $-1<n_{\sigma}(\pi)<2$, in jets with $\langle p_{\mathrm{T,jet}}\rangle\sim13.1~\mathrm{GeV}/c$.}
	\label{fig:app_PID_EC}
\end{figure}

\begin{figure}[htbp]
	\centering
	\includegraphics[width=1.0\linewidth]{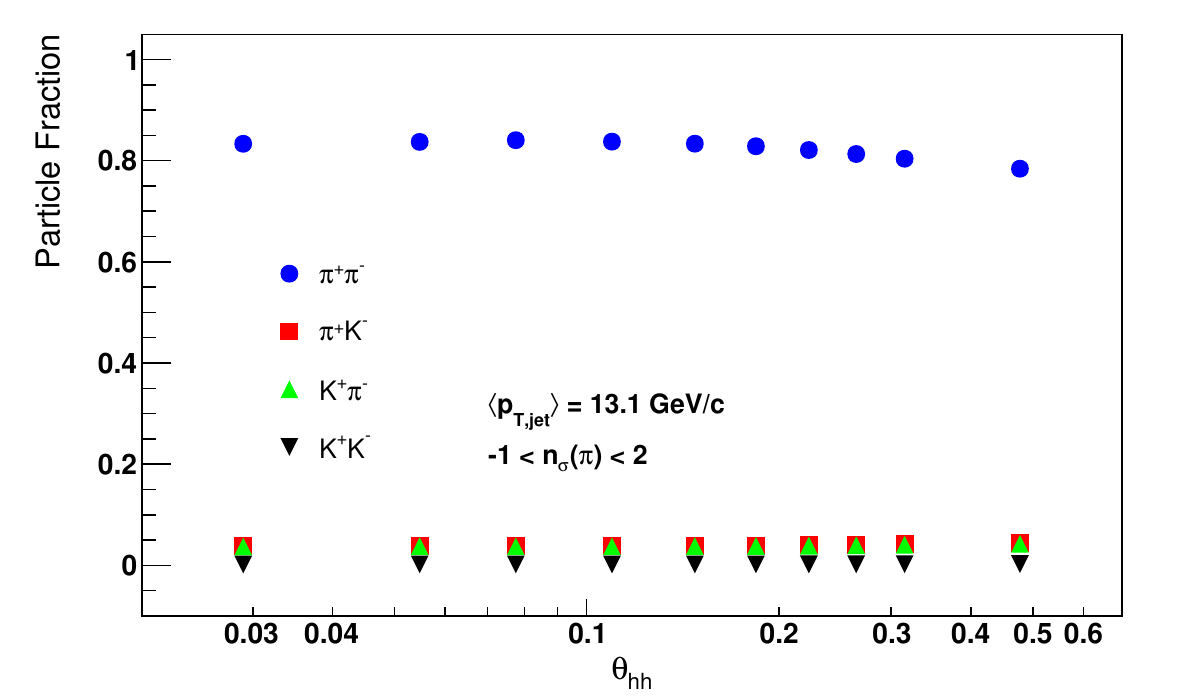}
    \caption{Fractions of charge-ordered particle pairs as a function of $\theta_{hh}$ for pairs in which both tracks satisfy the nominal pion-enriched selection, $-1<n_{\sigma}(\pi)<2$, in jets with $\langle p_{\mathrm{T,jet}}\rangle\sim13.1~\mathrm{GeV}/c$.}
	\label{fig:app_PID_EEC}
\end{figure}

The particle-identification (PID) fractions used in the purity unfolding are determined from fits to the TPC and TOF responses following the procedure of Ref.~\cite{STAR:2022hqg}. Figures~\ref{fig:app_PID_EC} and \ref{fig:app_PID_EEC} show representative single-track and charge-ordered pair fractions for the nominal pion-enriched selection, $-1<n_{\sigma}(\pi)<2$, in jets with $\langle p_{\mathrm{T,jet}}\rangle\sim13.1~\mathrm{GeV}/c$. The observed angular dependence of these fractions is incorporated bin by bin in the purity correction matrices used to unfold the measured one- and two-point asymmetries.

\begin{figure}[htbp]
	\centering
	\includegraphics[width=1.0\linewidth]{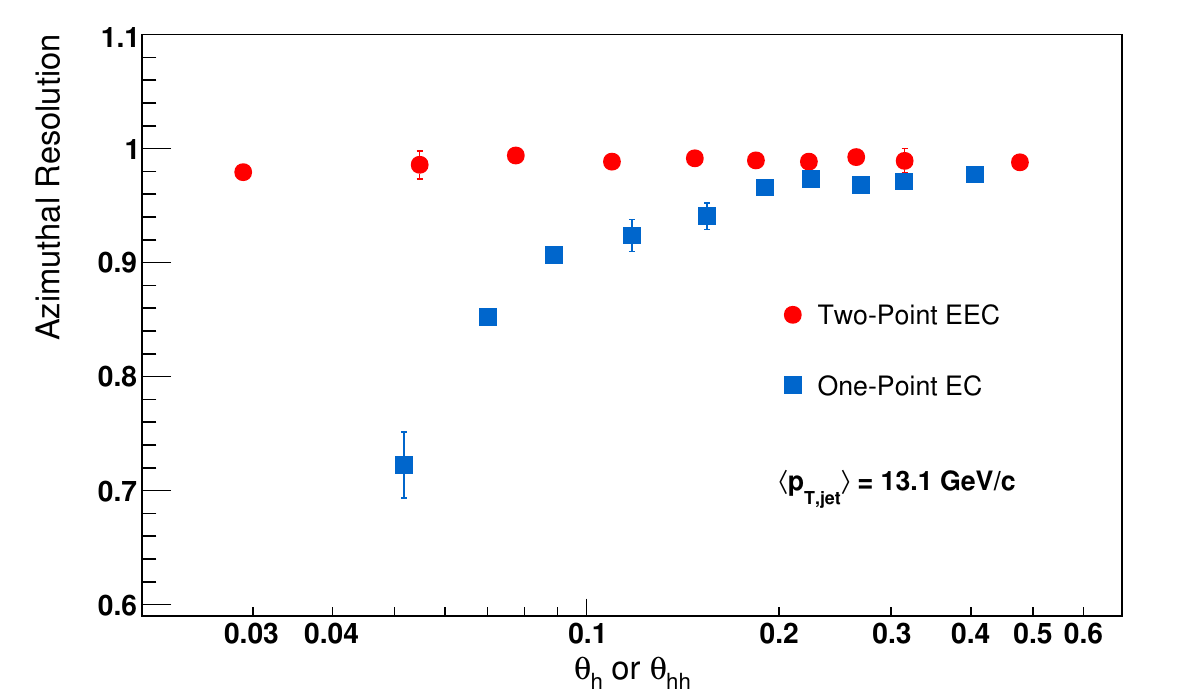}
    \caption{Resolutions of the azimuthal angle $\phi_S-\phi_h$ and $\phi_S-\phi_{hh}$ as functions of the corresponding angular distances for jets with $\langle p_{\mathrm{T,jet}}\rangle\sim13.1~\mathrm{GeV}/c$.}
	\label{fig:AngularRes}
\end{figure}

The azimuthal resolution is evaluated in simulation by comparing the reconstructed detector-level azimuthal angles with their matched particle-level values. Figure~\ref{fig:AngularRes} shows the representative resolutions of the one- and two-point azimuthal angle $\phi_S-\phi_h$ and $\phi_S-\phi_{hh}$ as functions of the corresponding angular distances for jets with $\langle p_{\mathrm{T,jet}}\rangle\sim13.1~\mathrm{GeV}/c$. These resolutions are used to correct the measured one- and two-point asymmetries.

The measured $\pi^+\pi^-$ asymmetries in each $\langle p_{\rm T,jet}\rangle$ interval were fitted with the empirical function
\begin{equation}
A_{\rm UT}(\theta_{hh}) =
C + A_0
\exp\left[
-\frac{\left(\ln\theta_{hh}-\mu\right)^2}{2\sigma^2}
\right].
\label{eq}
\end{equation}
The peak position and amplitude are given by
\begin{equation}
\theta_{hh}^{\rm peak}=e^\mu,
\qquad
A_{\rm UT}^{\rm peak}=C+A_0.
\end{equation}
The point-by-point statistical and systematic uncertainties were combined in quadrature in the fit.

\begin{table*}[t]
\caption{Empirical fit results for the two-point $\pi^+\pi^-$ asymmetry. The reference angle $\theta_{hh}^{\rm ref}$ is obtained by propagating the unpolarized transition scale from Ref.~[6].}
\begin{ruledtabular}
\label{app:peakfit}
\begin{tabular}{cccccc}
$\langle p_{\rm T,jet}\rangle$ (GeV/$c$) &
$\theta_{hh}^{\rm ref}$ &
$\theta_{hh}^{\rm peak}$ &
$\theta_{hh}^{\rm peak}/\theta_{hh}^{\rm ref}$ &
$A_{\rm UT}^{\rm peak}$ &
$\chi^2/{\rm ndf}$ \\
\hline
6.9 & $0.40\pm0.07$ & $0.25\pm0.02$ & $0.62\pm0.12$ & $0.027\pm0.011$ & 0.79 \\
8.9 & $0.31\pm0.06$ & $0.22\pm0.02$ & $0.71\pm0.15$ & $0.020\pm0.006$ & 0.76 \\
11.1 & $0.25\pm0.04$ & $0.23\pm0.02$ & $0.92\pm0.17$ & $0.026\pm0.004$ & 0.45 \\
13.1 & $0.21\pm0.04$ & $0.19\pm0.02$ & $0.90\pm0.20$ & $0.036\pm0.004$ & 0.36 \\
15.2 & $0.18\pm0.03$ & $0.10\pm0.03$ & $0.56\pm0.19$ & $0.043\pm0.005$ & 0.35 \\
19.6 & $0.14\pm0.02$ & $0.11\pm0.02$ & $0.79\pm0.18$ & $0.063\pm0.006$ & 0.56 \\
\end{tabular}
\end{ruledtabular}
\end{table*}

\begin{figure*}[t]
\centering
\includegraphics[width=\linewidth]{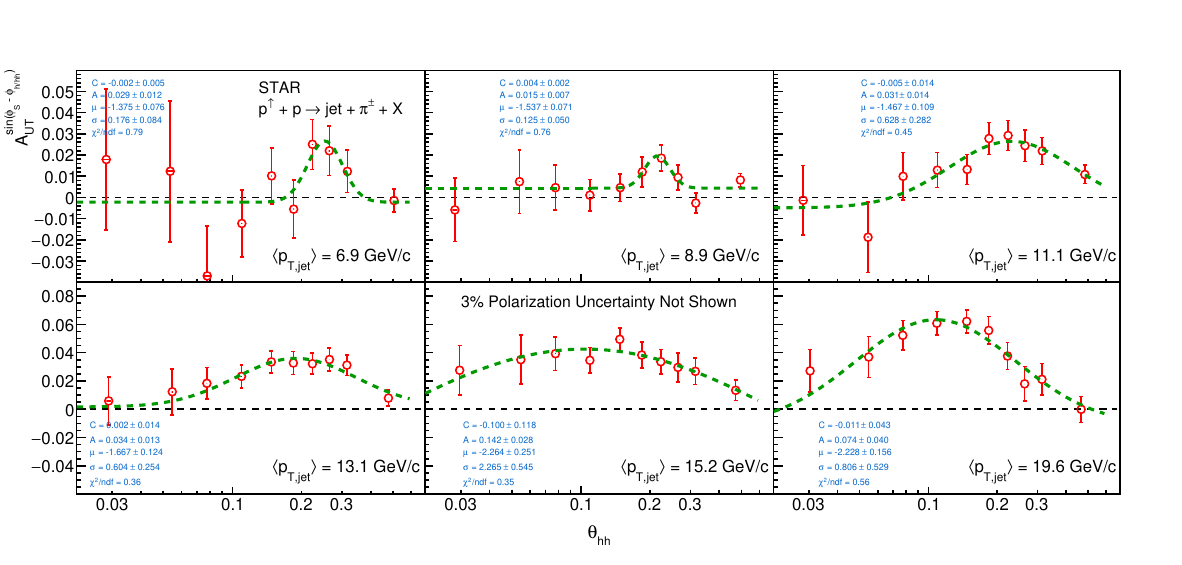}
\caption{Empirical fits of the two-point $\pi^+\pi^-$ transverse single-spin asymmetry in the six $\langle p_{\rm T,jet}\rangle$ regions. Curves show the log-Gaussian parameterization of Eq.~(\ref{eq}). The fitted peak positions are summarized in Table~\ref{app:peakfit}.}
\label{fig:EEC_PeakFit}
\end{figure*}

Figure~\ref{fig:EEC_PeakFit} shows the fits, and the corresponding comparisons are presented in Tab.~\ref{app:peakfit}. 

\end{document}